\documentclass[aps,pra,twocolumn,showpacs,groupedaddress]{revtex4}  
\usepackage{graphicx}  
\usepackage{dcolumn}   
\usepackage{bm}        
\usepackage{amssymb}   
\usepackage{verbatim}  

\newcommand{\abs}[1]{\ensuremath{\left| #1 \right|}}

\newcommand{\beq}{\begin{equation}}
\newcommand{\eeq}{\end{equation}}

\begin{document}


\hspace{5.2in} \mbox{Fermilab-Pub-04/xxx-E}

\title{Classical and quantum dynamics of pulsating instability of a Bose-Einstein condensate in an optical lattice}
\author{Uttam Shrestha}
\affiliation{Department of Physics, University of Connecticut,
Storrs, CT 06269-3046}
\date{\today}

\begin{abstract}
We study the dynamics of a Bose-Einstein condensate (BEC) in a one dimensional optical lattice in the limit of weak atom-atom interactions by incorporating quantum fluctuations. The pulsating dynamical instability manifests itself in the time evolution in which atoms periodically collect themselves into a pulse and subsequently disperse back into the initial homogeneous state. We take into account the quantum fluctuations within truncated Wigner approximation and observe that the quasiperiodic behavior still persists for single realizations which may represent the typical experimental outcome. The quantum mechanical ensemble averages of the wave functions shows a damping in the pulsating event. The fluctuations become more prominent for smaller atom numbers.  

\end{abstract}

\maketitle

\section{\label{sec:level1}Introduction}

The superfluidity of a Bose-Einstein condensate (BEC) in an optical lattice has been drawing a considerable attention in last several years \cite{Morsch}. As is well known, superflow of the BEC in free space suffers from an instability when the center of mass velocity reaches a critical value. Such an instability, known as Landau or energetic instability, exist when the superfluid flow is not at a local minimum of energy and the system lowers its energy by emitting phonons \cite{Pethick}. In an optical lattice, in addition to the energetic instability, the BEC may also exhibit dynamical or modulational instabilities which have been a subject of active experimental and theoretical research in recent years \cite{BUR01,Wu,SME02,CAT03,CRI04,FAL04,SAR05,WU03, Trombettoni,FER05,ZHE04,MON04,RUO05,BAN06,RUO07,Barontini,Ferris,Shrestha}. When the system is in the dynamically unstable regime, small perturbations grow exponentially in time resulting in an irregular dynamics, loss of coherence or an abrupt stop of the transport of the atom cloud \cite{Wu, SME02}.

In this paper we study dynamical instabilities of atoms in an optical lattice for the case of weak atom-atom interactions and also taking into account quantum fluctuations of atoms. We recently reported \cite{Shrestha} that, by appropriately selecting the strength of the interactions, the corresponding classical system may exhibit a pulsating dynamical instability in which the atoms nearly periodically collect to a peak in lattice occupation numbers, and subsequently disperse back to (very close to) the initial unstable state. This is different from the conventional view, valid at strong interatomic interactions, that dynamical instabilities for BECs in optical lattices are associated with irregular dynamics. When we incorporate quantum fluctuations of atoms using stochastic phase-space methods, the quasiperiodic behavior is still observable in individual stochastic realizations that represent typical individual experimental realizations. As the pulsating solitons in each realization appear at different lattice sites due to quantum effects, the quantum mechanical ensemble averages of the wavefunction revival become progressively weaker when the effective interaction strength is increased. Other ensemble averages, such as the pulsation amplitude, can still provide information about the quantum soliton.

We consider a stationary superfluid flow of a BEC in an optical lattice with a large enough flow momentum that triggers the dynamical instability of the corresponding classical nonlinear system. In a quantum system the corresponding sharp transition to the dynamically unstable regime is smeared out, typically resulting in a progressively increasing dissipation in the dynamics close to the classical onset of the instability \cite{RUO05}. We provide a qualitative explanation of the pulsating phenomenon by studying the dynamics of an integrable double-well system. Although the instability is a result of the interplay between the lattice discreteness and the nonlinearity that makes the lattice non-integrable, the dynamics of the lattice with many sites is approximately as if the system is integrable. Related classical pulsations starting from already compressed atom distribution in a lattice have been discussed in \cite{Barontini} within the frame-work of the nonpolynomial Schr\"odinger equation.


The pulsating instability manifests in the dynamical regime where the nonlinearity is weak. In the mean-field description, the size of the nonlinearity  is proportional to the atom-atom interactions and the total number of atoms present in the system. As the number of atoms gets small the mean-field description may breakdown as the relative fluctuations in the system amplifies, and the quantum treatment is inevitable. We  would like to know how the quantum effects smear out the pulsating mechanism as we reduce the number of atoms in the system. At the simplest level, we study the quantum dynamics of the pulsating instability using the quantum distribution function, in particular, the Wigner function method. The Wigner method simulates the quantum mechanical system in classical stochastic process where the quantum fluctuation is included  in the initial state. In the case of BEC it gives the time evolution of the whole matter field including both condensate and non-condensate atoms, and allows the scattering between them, which is absent in the classical GP description.

In Sec. II we formulate the theoretical model, mainly the Gross-Pitaevskii equation (GPE) \cite {Gross} and its discrete variant, the discrete nonlinear Schr$\ddot{\text o}$dinger equation (DNLSE) \cite{SME02, Christ}. We use linear stability analysis to find the region of interaction strengths and flow quasimomenta  where the system develops instability. As in nonlinear dynamics, following \cite{Hennig}, we verify the existence of the localized soliton solution in the forbidden gap of the linear spectrum. In Sec. III we investigate the time evolution of the DNLSE within classical mean-field theory. Although the system initially develops instability the time evolution shows a regular dynamics whereupon the atoms periodically collect themselves into a pulse and disperse back into the unstable state.

In Sec. IV we review the well understood double well system and argue that the dynamical behavior of the multi-site system is analogous to the two-site system, at least in the limit of weak nonlinearity. In Sec. V we study the dynamics beyond the classical mean field theory using truncated Wigner approximation (TWA); a phase-space method that approximately solves the dynamics of a quantum system by means of stochastic initial configuration. We then compare various physical properties such as the number fluctuations and the overlaps of the state of the system in  single realizations with an ensemble averages. Quantum dynamics significantly modifies the classical picture as the number of particles gets small. We observe the damping in the pulsating phenomenon when we average over many stochastic trajectories.

\section{\label{sec:level1}Theoretical Model:  DNLSE}

At absolute zero temperature the dynamics of the BEC atoms in an optical lattice can be modeled by the mean field Gross-Pitaevskii equation \cite{Gross,Pethick}

\begin{eqnarray}
i\hbar {d\Psi\over dt}
=\left(-{\hbar^2\over 2 m}{\Delta}+V({\bf x})+g_{3D}|\Psi|^2 \right ) \Psi,
\label{ee1}
\end{eqnarray}
where $\Psi({\bf x},t)$ is a wave function corresponding to the bosonic field operator such that $|\Psi({\bf x})|^2 =n({\bf x})$, the atom density.
The coupling constant $g_{3D}$  is related to the scattering length through $g_{3D}=\frac{4\pi \hbar^2 a_s}{m}$, where $a_s$ and $m$ are the s-wave scattering length and atomic mass respectively. The positive and negative  scattering lengths respectively correspond to the repulsive and attractive atom-atom interactions. The Eq. (\ref {ee1}) is an approximate description of an assembly of a large number of bosonic atoms that are in the same quantum mechanical state. 

We consider the external potential, $ V({\bf x})$ of the form
\begin{equation}
V({\bf x})={1\over 2} m (\omega_x ^2 x^2+\omega_{\perp} ^2 r_{\perp}^2)+V_0 \sin^2\left(\frac{\pi x}{d_L}\right).
\label{e2}
\end{equation}
Here $V_0$  and $d_L(=\lambda/2)$  are respectively the depth and the periodicity of the optical lattice. If the harmonic confinement is much stronger in the transverse than in the longitudinal direction $(\omega_{\perp}\gg \omega_x)$ the GPE can be transformed into a one-dimensional form

\begin{eqnarray}
i\hbar {d\psi\over dt}
=\left(-{\hbar^2\over 2 m}{\frac{\partial^2}{\partial x^2}}+V({x})+g|\psi|^2 \right ) \psi,
\label{e1}
\end{eqnarray}
with an effective atom-atom interactions $g=2 a_s \hbar \omega_{\perp}$.
When the depth of the optical lattice is much larger than the chemical potential of the atoms, one can employ the tight-binding approximation. By expressing the condensate wave function $\psi(x)$ as a superposition of the Wannier functions localized within each potential well of the lattice, one can obtain the tight-binding version of the GPE known as the discrete nonlinear Schr$\ddot{\text o}$dinger equation (DNLSE) \cite{SME02}:
\begin{equation}
i{\partial\over\partial t}\psi_{n}=-J(\psi_{n+1}+\psi_{n-1})+(V_{n}+\chi \abs{\psi_{n}}^2)\psi_{n}.
\label{e3}
\end{equation}

The parameters $J$ and $V_n$ respectively characterize the tunneling rate and the external trapping potential, whereas $\chi$ is proportional to the atom-atom interactions.
It is convenient to scale the wave function and interaction parameter as $ \psi_n \equiv \sqrt{N_T}\psi_n $ and $\chi\equiv N_T \chi$, with $N_T$ being the total atom number, so that the wave function is properly normalized to one. Unless it is explicitly stated otherwise, we assume here that $J>0$ and the atom-atom interaction is repulsive, $\chi\geq 0$.

In the absence of external potential and nonlinearity  the Eq. (\ref {e3}) may be solved with a plane-wave ansatz $\psi_n(t)=\frac{1}{\sqrt{N}}\text e^{i(pn- \omega(p) t)}$, giving the dispersion relation $\omega(p)=-2 J\cos p$. The periodic boundary conditions quantize the quasimomenta  $p=2\pi P/N$, $N$ being  the number of lattice sites, and $P$ is an integer that may be chosen to lie in the interval $[-\frac{N}{2},\frac{N}{2})$. For notational convenience we always take the number of lattice sites $N$ to be even.

When the interaction is switched on, the constant-amplitude plane waves are still solutions to Eq. (\ref{e3}) but with a modified dispersion relation $\omega(p)=-2 J\cos p+ \frac{\chi}{N}$. Besides these extended-wave solutions, the nonlinear system also admits solutions that are localized in space \cite{Dauxois}. These solutions, so called the gap solitons,  usually have an energy that lies outside of the linear band spectrum. In the continuum model solitonic solutions of the nonlinear Shr$\ddot {\text o}$dinger equation can be found in closed form by using the inverse scattering method \cite{Akhmediev}. However, the discrete system has fewer constants of the motion and is not integrable as such, so that one has to rely on numerical techniques.

\subsection{\label{sec:level1}Hamiltonian and its symplectic nature}

The Hamiltonian corresponding to the equation of motion Eq. (\ref{e3}) is given by
\begin{equation}
 H=\sum_n\bigg{\{}-J(\psi_n\psi_{n+1}^*+h.c.)+V_n|\psi_n|^2+{\chi\over 2}|\psi_n|^4\bigg{\}},
\label{e7}
\end{equation}
where $\psi_n$ and $~i\psi_n^*$ are canonically conjugate variables that satisfy Hamilton's equations of motion
\begin{equation}
\psi_n=-{\partial H\over \partial (i\psi_n^*)},\quad i\psi_n^*={\partial H\over \partial \psi_n}.
\label{e8}
\end{equation}
Although we are dealing with a system with a quantum origin, the macroscopic wave function, nonetheless, obeys classical equations of motion. Since the time is cyclic in the Hamiltonian,  the total energy is a constant of the motion. The normalization which is proportional to the total number of particles, is also a constant of the motion.

To study the basic features of the solutions governed by the Hamiltonian (Eq. (\ref{e7})) near the edge of the linear band spectrum, we consider Eq. (\ref{e3}) as a map where the lattice indices play the role of the discrete time \cite{Hennig}. Without the loss of generality and for simplicity in our continuing discussion in the following we neglect the effect of external trapping potential so that the system is translational invariant along the lattice direction. First we write down the stationary state solution of the Hamiltonian in the form $\psi_n(t)=\psi_n \exp(-i \omega t)$ to obtain the time independent equation
\begin{equation}
\omega \psi_{n}=-J(\psi_{n+1}+\psi_{n-1})+\chi \abs{\psi_{n}}^2\psi_{n},
\label{ee3}
\end{equation}
and then separate the real and imaginary parts, $\psi_n=x_n+i y_n$, resulting an area preserving 4-dimensional real map $\cal {\bf M:}$
\begin{eqnarray}
     x_{n+1} &=& \chi(x_n^2+y_n^2)x_n-u_n-\omega x_n,\nonumber\\
	y_{n+1} &=& \chi(x_n^2+y_n^2)y_n-v_n-\omega y_n, \nonumber\\
	u_{n+1} &=& x_n,\nonumber\\
	v_{n+1} &=& y_n.
\label{e9}
\end{eqnarray}
Here, for convenience, we take $J=1$. Given the initial conditions $(x_0,y_0,u_0,v_0)$, one can propagate the solution for a given energy $\omega$ to obtain an orbit of the discrete lattice system.
The Jacobian matrix of the map $\cal{\bf M}$ is given by

\[\left( \begin{array}{cccc}
	a & b & -1 & 0 \\
	b & c & 0 & -1  \\
	1 & 0 & 0 & 0  \\
	0 & 1 & 0 & 0
	\end{array}\right)\]
where
\begin{eqnarray*}
 a&=&\gamma+2\chi x_n^2,\\
 b&=&2 \chi x_n y_n,\\
 c&=&\gamma+2 \chi y_n^2,\quad \text{and} \\
 \gamma&=&\chi(x_n^2+y_n^2)-\omega.
\end{eqnarray*}
Since the determinant of the Jacobian matrix is  one, the map is indeed area preserving \cite{Ott}. The fixed point of the map is (0,0,0,0). In order to study the stability of the fixed point one has to solve for the roots of the \emph{characteristic polynomial}
\begin{equation}
(\lambda (\omega+\lambda)+1)^2=0,
\label{e10}
\end{equation}
which gives the corresponding eigenvalues
\begin{equation}
\lambda_{\pm}={-\omega\pm (\omega^2 -4)^{1/2}\over 2}.
\label{e11}
\end{equation}
Since $\lambda_{+} \lambda_{-}=1$, the roots are reciprocal of each other indicating the symplectic nature of the Hamiltonian. There are three possibilities:\newline
(a) $-2<\omega<2$; all the roots are complex with magnitude one;\newline
(b) $\omega>2$; the roots are real and negative;\newline
(c) $\omega<-2$; the roots are real and positive.\newline
In case (a) the periodic orbit corresponding to the fixed point $(0,0,0,0)$ is elliptical and is stable. In case (b) and (c) the periodic orbits are hyperbolic and is unstable.
\begin{figure}
\includegraphics[clip,width=1.0\linewidth]{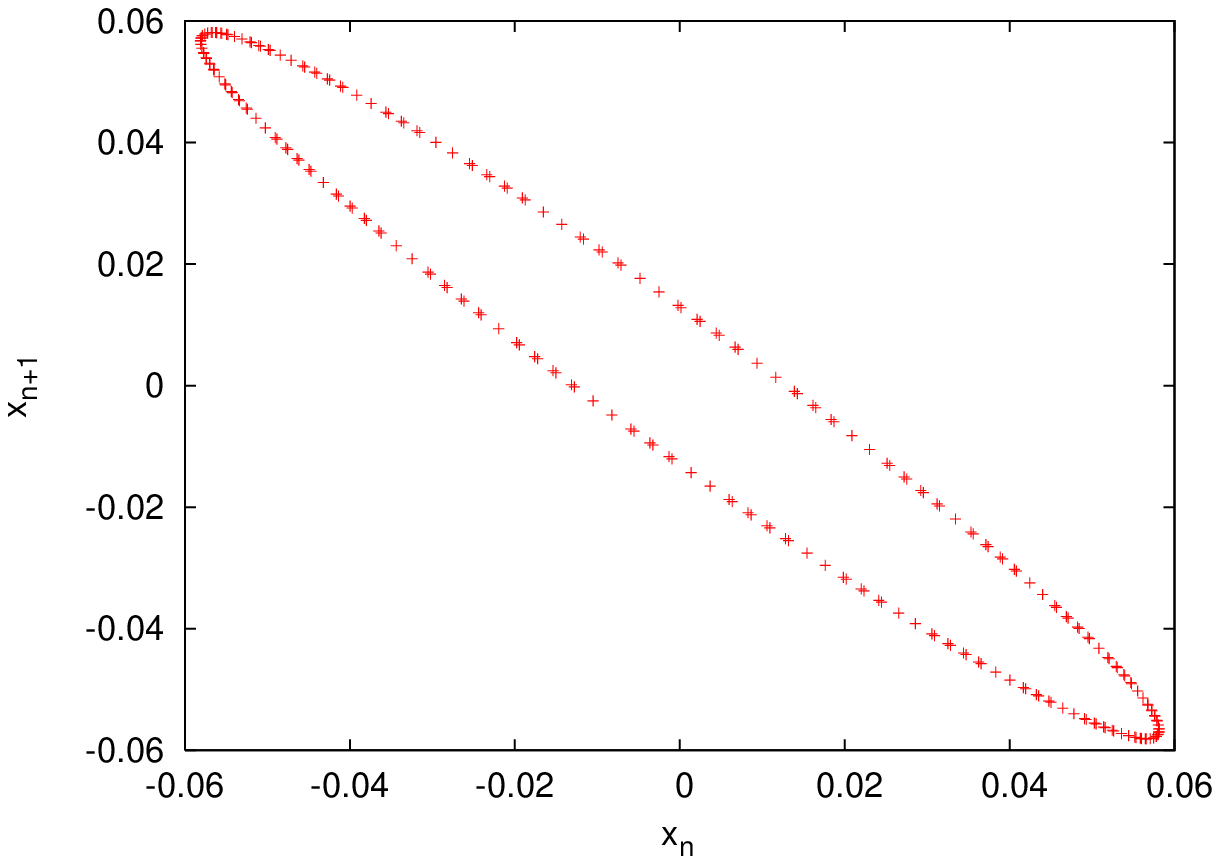}
\includegraphics[clip,width=1.0\linewidth]{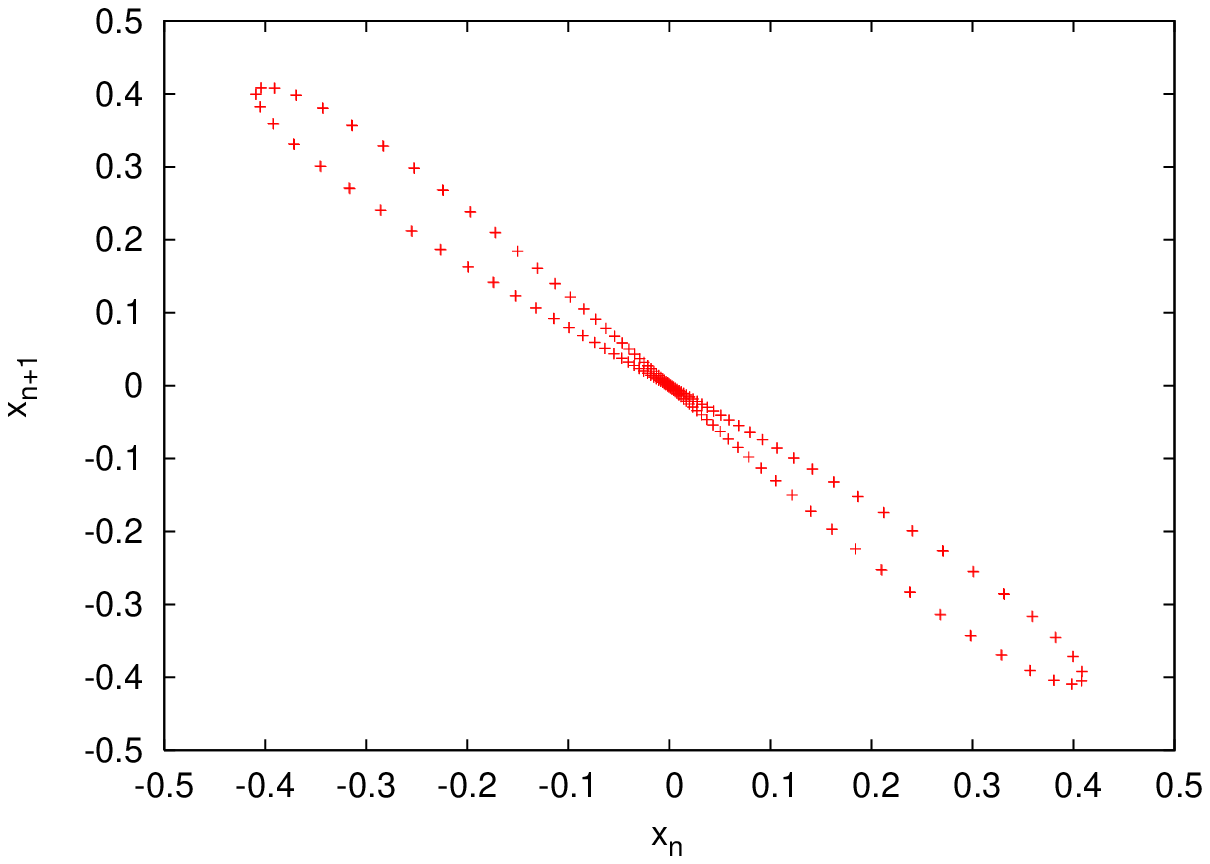}
\caption{Map orbit around an elliptical fixed point for two energies $\omega=1.95$ (top) and $2.05$ (bottom). The stable elliptical fixed point bifurcates leading to the period doubling bifurcation when the energy crosses through the critical point. The changing of the elliptical fixed point into hyperbolic is an indication of the emergence of different type of solution.}
\label{f1}
\end{figure}
In Fig. \ref {f1} we have plotted an orbit of the map $\cal {\bf M}$ for $\omega=1.95$ (top) and $\omega=2.05$ (bottom). It is clearly seen that the elliptical fixed point $(0,0,0,0)$ loses its stability and turns into an unstable hyperbolic point with the onset of period doubling bifurcation when $\omega$ passes through the critical value two. This hyperbolic fixed point lies on a homoclinic orbit that corresponds to spatially localized soliton solution. Furthermore, it should be noted that Eq. (\ref {ee3}) is invariant under transformation $\psi_n\rightarrow (-1)^n \psi_n,~ \chi\rightarrow -\chi,~ \omega\rightarrow -\omega$, every solution in the positive region of the linear band spectrum has one to one correspondence to the negative one. We, thus, expect soliton solutions in a lattice for both repulsive and attractive atom-atom interactions. However, these two solutions differ intrinsically in the sense that the solitons in the repulsive case have an alternating signs between adjacent lattice sites whereas in the attractive case, they have the same sign.

\subsection{\label{sec:level1}Modulational Instability}

The DNLSE admits stationary solutions of the form  $\psi_n(t)=\frac{1}{\sqrt{N}}\text e^{i(p n-\omega(p)t)}$ with the dispersion $\omega(p)=-2J\cos p+\frac{\chi}{N}$. To study the stability of a solution we introduce an infinitesimal perturbation around the steady state \cite {Wu,SME02},
\begin{equation}
\psi_n(t)=\psi_n^0(t)[1+u e^{i(q n-\Omega t)}+v^*e^{-i(q n-\Omega^* t)}],
\label{e12}
\end{equation}
where $q$ and $\Omega$ are the momentum and the frequency of the small excitation relative to the initial unperturbed steady state solution. After inserting Eq. (\ref {e12}) in Eq. (\ref {e3}) with $V_n=0$, and expanding to the lowest nontrivial order in $u$ and $v$ we get the following matrix equation,
\begin{equation}
i{d{\xi}\over dt}={\cal M}{\xi},
\label{e13}
\end{equation}
where $\xi$ is a vector $[u,v]^T$ and ${\cal M}$ is a $2\times 2$ matrix with elements
\begin{eqnarray}
&&{\cal M}_{11}=\frac{\chi}{N} +4 \sin \frac{q}{2}~\sin(\frac{q}{2}+p),\nonumber\\
&&{\cal M}_{22}=-\frac{\chi}{N} -4 \sin \frac{q}{2}~\sin(\frac{q}{2}-p),\nonumber\\
&&{\cal M}_{12}=-{\cal M}_{21}^q=\frac{\chi}{N}.	
\label{e14}
\end{eqnarray}
The eigenvalues of $\cal M$ give the small-excitation frequencies,
\begin{eqnarray}
&&\Omega= 2J \sin p \sin q\nonumber\\
&&\pm\sqrt{4 J \cos p(1-\cos q)[ \frac{\chi}{N} +J \cos p (1-\cos q)]},
\label{e15}
\end{eqnarray}
whereas the eigenvectors give the corresponding mode functions:
\begin{equation}
u=\sqrt{\frac{\Omega-{\cal M}_{22}}{2\Omega-{\cal M}_{11}-{\cal M}_{22}}},
\label{ee16}
\end{equation}
\begin{equation}
v=\sqrt{-\frac{\Omega-{\cal M}_{11}}{2\Omega-{\cal M}_{11}-{\cal M}_{22}}}.
\label{ee17}
\end{equation}
Substituting Eqs. (\ref {e14}), (\ref{e15}) into Eqs. (\ref {ee16}), (\ref{ee17}), we obtain

\begin{equation}
u=\sqrt{\frac{4 J \cos (p) \sin ^2\left(\frac{q}{2}\right)+\Gamma +\frac{\chi}{N}
   }{2 \Gamma }},
\label{ee18}
\end{equation}
\begin{equation}
v=\sqrt{\frac{4 J \cos (p) \sin ^2\left(\frac{q}{2}\right)-\Gamma +\frac{\chi}{N}
   }{2 \Gamma }},
\label{ee19}
\end{equation}
with the definition
\begin{eqnarray}
\Gamma=
\pm\sqrt{4 J \cos p(1-\cos q)[ \frac{\chi}{N} +J \cos p (1-\cos q)]}.
\label{ee20}
\end{eqnarray}
By inspection it can be easily verified that the mode functions $u$ and $v$ satisfy the normalization condition $|u|^2-|v|^2=\pm 1$ as long as $\Gamma$ is real.

The eigenvalues $\Omega$ corresponding to positive normalization gives the physical small-excitation frequencies whereas that corresponding to negative normalization are unphysical. Since we are dealing with the repulsive atom-atom interactions, $\chi\geq 0$, it is seen that all eigenvalues are real for $|p|\leq{\pi\over 2}$. However, the existence of complex eigenvalues cannot be ruled out in the interval $|p|\in [\frac{\pi}{2},\pi]$, depending on the values of $J,~\chi,~q$ and $p$. When an eigenvalue is complex, i.e., $\Gamma$ is imaginary, small perturbations in the steady flow grow exponentially in time. In this case the norm of the eigenvector  $|u|^2-|v|^2$ vanishes identically \cite{Wu}.

For any $|p|$ greater than ${\pi\over 2}$ and for large $N$ , the eigenfrequencies will be complex if
\begin{equation}
\Lambda>|\cos(p)|~\frac{\pi^2 Q^2}{N},
\label{e16}
\end{equation}
where $Q$ is a non-zero integer that lies in the interval $[-\frac{N}{2},\frac{N}{2})$ and the rescaled interaction strengths $\Lambda$ is defined to be 
\begin{equation}
\Lambda=\frac{\chi}{2J}.
\label {ep1} 
\end{equation}
The flow with the quasimomentum $p$ is then said to be dynamically unstable in the sense that a small noise drives the system far away from the equilibrium state. It should be pointed out that the critical interaction strength approaches zero when the number of lattice sites goes to infinity, implying that any flow with $|p|>{\pi\over 2}$ and a fixed $\Lambda>0$ will turn unstable with $N\rightarrow \infty$. The dynamical instability can be qualitatively understood from the dispersion relation of the DNLSE. For $|p|>{\pi\over 2}$, equivalently when the effective mass is negative, the interaction shifts the frequency upward in the forbidden gap of the linear spectrum where the plane wave solution cannot exist, which means that the system is unstable. Moreover, the imaginary part of the complex eigenfrequency as well as the corresponding eigenvectors are the same for $q$ and $-q$ (see Eqs. (\ref{ee18}), (\ref{ee19})), which indicates that these two modes are equivalent as it comes to the instability.

\begin{figure}
\includegraphics[clip,width=1.0\linewidth]{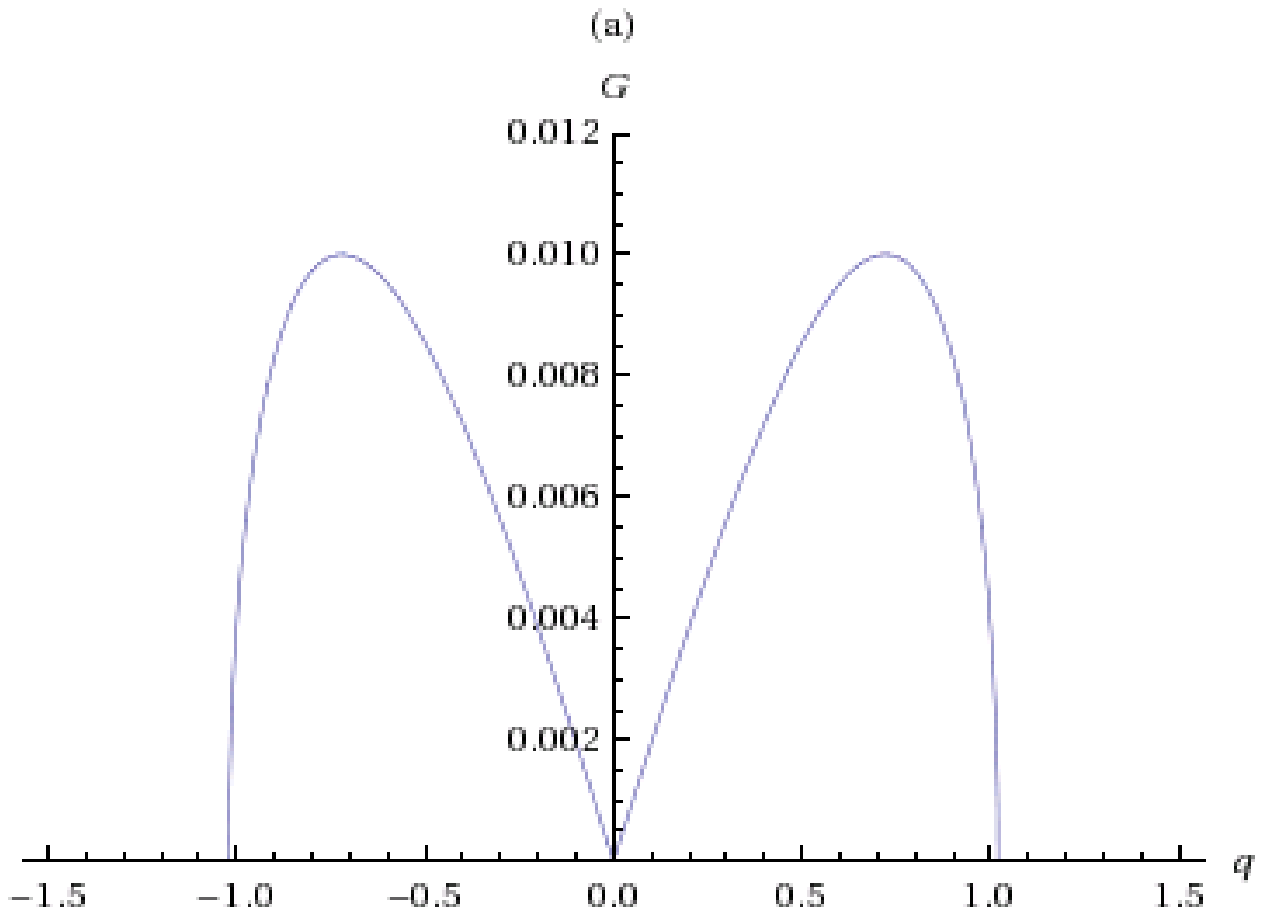}
\includegraphics[clip,width=1.0\linewidth]{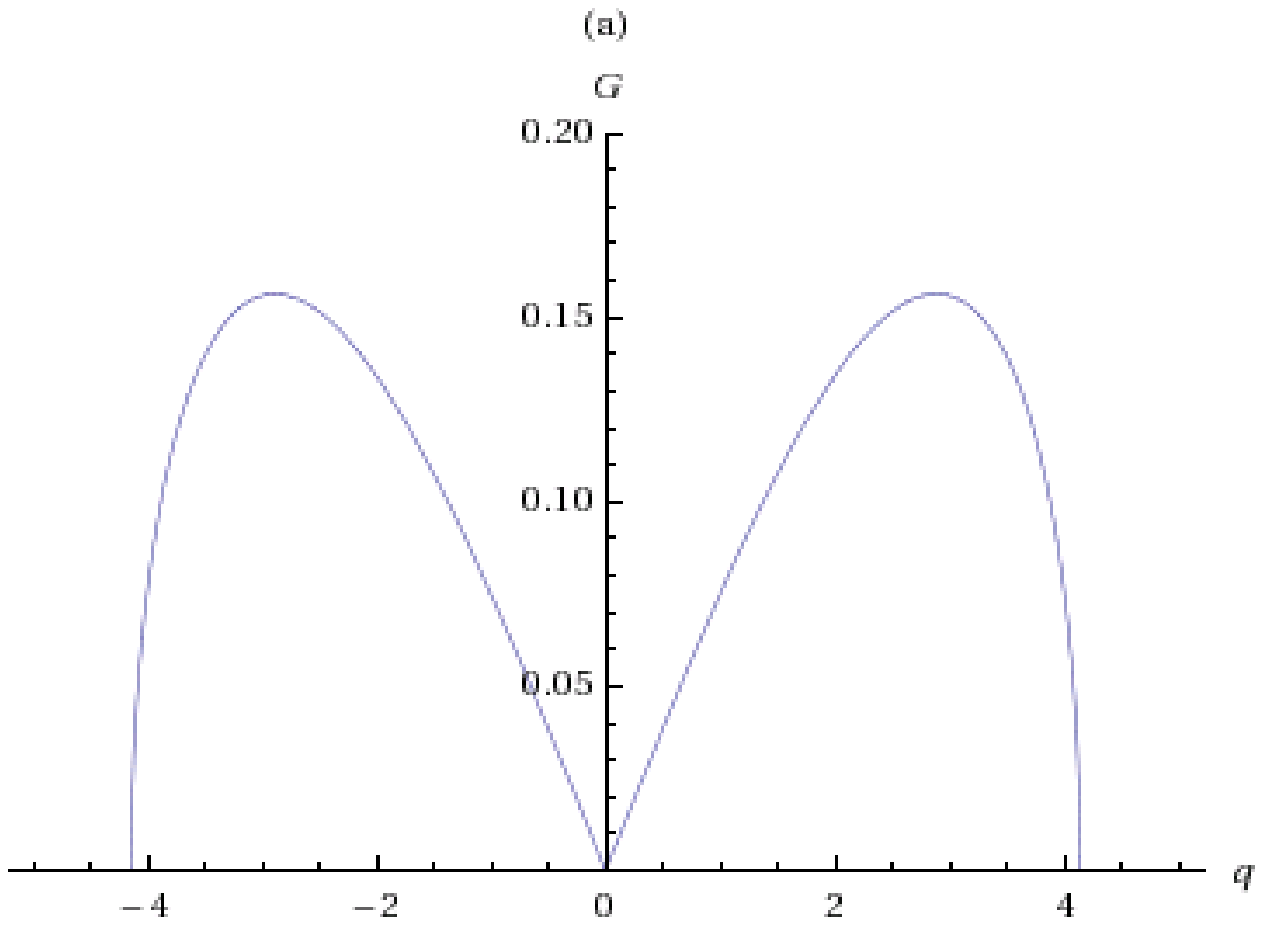}
\caption{Coefficient of exponential gain $G$ for an excitation mode with quasimomentum $q$ for the interaction strengths $\Lambda=0.32$ (a) and $\Lambda=5.0$ (b). Modulational instability is only possible for $G>0$, and excitation modes may only occur for nonzero integer values $Q$ of $q$.}
\label{f2}
\end{figure}

Fig. \ref{f2} shows the gain curve, $G=\text{min}[0, |\text{Im}~ \Omega(p,q)|]$, for two interaction strengths (a) $\Lambda=0.32$, and (b) $\Lambda=5.0$ in a lattice of $32$ sites with $p=\pi$ for $-\infty \leq q \leq \infty$. The figure (a) reveals a single pair of sidebands with one unstable mode, whereas the figure (b) shows four pairs for four unstable modes.

\section{\label{sec:level1}Time Evolution and Pulsating Instability}

We carry out numerical simulations on the DNLSE to study the growth of the unstable mode in a lattice for a suitable range of interaction parameters. For a given number of lattice sites  and the flow momentum the number of unstable modes in the linear stability analysis depends only on the interaction parameter $\Lambda$. Here we focus only on low energy excitations in the limit of weak atom-atom interactions. Two numerical methods have been used for the time evolution, an unconditionally stable Crank-Nicholson type algorithm \cite{NR} and a sixth-order accurate FFT split operator algorithm that works in the same way as is discussed in \cite{Javanainen} for the ordinary nonlinear Schr$\ddot{\text o}$dinger equation.

\subsection{\label{sec:level1}Single Unstable Mode}

A straightforward analysis of the eigenfrequency expression Eq. (\ref{e15}) suggests that the range of interaction strengths where the $Q=1$ mode is unstable but the $Q=2$ mode is not is given by

\begin{equation}
 |\cos p|~\frac{\pi^2}{N} < \Lambda < 4 |\cos p|~\frac{\pi^2}{N}.
\label{e17}
\end{equation}

\begin{figure}
\includegraphics[clip,width=1.0\linewidth]{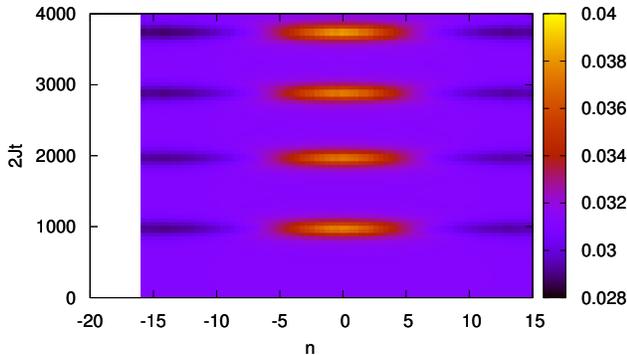}
\caption{Collapse and revival of the density hump in the BEC evolution when the lattice starts from a dynamically unstable state. The parameters are $\Lambda=0.48$, $p=\pi$, and $N=32$. These parameters correspond to a single unstable mode, as per linear stability analysis. The instability drives the initial homogeneous atom distributions into a density peak that subsequently disperses back to the initial state.}
\label{f3}
\end{figure}

Fig. \ref{f3} shows a typical density plot of the time evolution of the BEC initially prepared in the plane wave state at the edge of the Brillouin zone ($p=\pi$) seeded with random Gaussian noise, for the number of lattice sites $N=32$ and the interaction strength $\Lambda=0.48$. This value of $\Lambda$ corresponds to one unstable mode in the linear stability analysis. Although a tiniest amount of noise (either in real experiments or in numerical simulations) in the unstable direction triggers the instability, an external noise of amplitude $\xi=10^{-4}$ is added just to speed up the instability. It has been tested in a number of runs that the time for the  onset of the instability for fixed values of the other parameters depends logarithmically on the amplitude of the added noise.

\begin{figure}
\includegraphics[clip,width=1.0\linewidth]{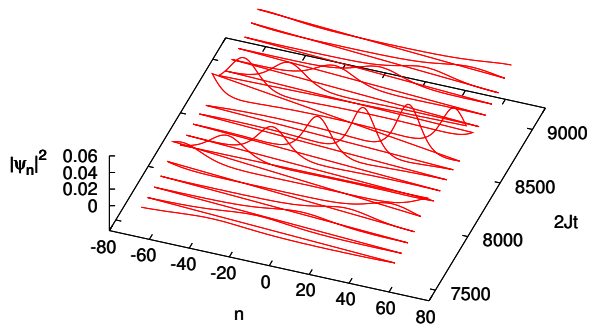}
\caption{Density evolution of the BEC for the initial flow momentum different from $\pi$. Here the parameters are $N=128,~\Lambda=0.25$ and $p=\frac{15~\pi}{16}$. The pulsating behavior of the peak still persists but the peak moves with the group velocity $v_g=\sin(p)$. }
\label{f4}
\end{figure}

Fig. \ref{f4} depicts a snapshot of a pulse that moves during its formation from the initial flow state with quasimomentum $p=\frac{15~\pi}{16}$. In this figure we take a larger lattice with $N=128$ sites and the interaction strengths is $\Lambda=0.25$, so that there is still one and only one unstable mode. By virtue of the periodic boundary conditions a pulse that goes over the right edge will reappear at the left edge of the lattice.

\begin{figure}
\includegraphics[clip,width=1.0\linewidth]{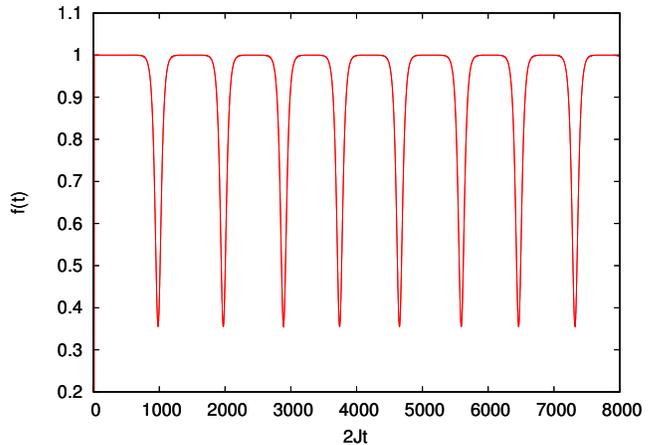}
\caption{Fraction of the initial state $f(t)$ in the state of the lattice plotted as a function of time. Here the parameters are $N=32, ~\Lambda=0.48$ and $p=\pi $.}
\label{f5}
\end{figure}

The pulsating behavior of the peak can also be viewed by plotting the fraction of the initial state $\psi_n(0)$ remaining in the state of the lattice  as a function of time,
\begin{equation}
f(t)=|\sum_n \psi_n ^*(0) \psi_n(t)|^2.
\end{equation}
In Fig. \ref{f5} we plot the overlap, $f(t)$, as a function of time for the same parameters as in Fig. \ref{f3}. It is revealed that the instability drives the system far from, and subsequently brings it back to, the original unstable steady state, and the process repeats. Each dip in the plot represents formation of a pulse during the course of time. It is also noted that the quantity $f(t)$ does not vanish all the way to zero, indicating that the pulsed state is not orthogonal to the initial steady state. Furthermore, a closer inspection of this plot shows that the subsequent peaking events are not strictly periodic; the interval between the dips varies slightly, implying a quasi-periodic phenomenon.

By analyzing data sets of this kind a number of interesting observations emerges. (i) First, in contradiction to the common belief that the instability may develop an irregular dynamics, it causes the atoms to pile up in a single-peaked distribution of the occupation numbers $|\psi_n|^2$. However, upon further time evolution, the system returns very close to the initial unstable state, again pulsates to a peak, and so on. We have periodic peaking and recurrences to the unstable initial state. (ii) Second, the peak may occur at any lattice site. It is the random noise that seeds the position of the peak. In order to test this claim, we ran the simulation a number of times with everything else except the particular realization of the noise held unchanged, and observed that peaking occurs approximately at the same time but the position of the peak is completely random. This is in accordance with the theory of the translational invariance of the lattice: A lattice-translated pulsed solution is also a degenerate solution of the DNLSE and there is no preferable lattice site for the occurrence of the pulse.
(iii) Third, the periodic recurrences and the velocity of the peak for a given number of lattice sites, seem to depend on the values of interaction strengths $\Lambda$ and initial flow momentum $p$ only. For the initial flow state $p\neq \pi$, the pulse moves with the velocity that turns out to be the group velocity of the carrier wave, $v_g =\sin(p)$.

\subsection{\label{sec:level1} Multiple Unstable Modes}

Equation (\ref{e17}) can be generalized to obtain the conditions for $Q$ unstable modes,
\begin{equation}
 Q^2|\cos p|~\frac{\pi^2}{N} < \Lambda < (Q+1)^2|\cos p|~\frac{\pi^2}{N}.
\label{e18}
\end{equation}

\begin{figure}
\includegraphics[clip,width=1.0\linewidth]{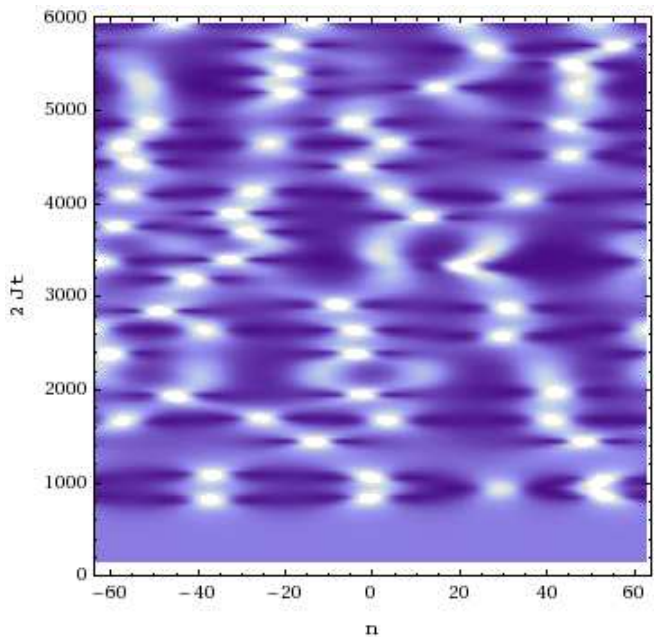}
\caption{Density plot of the populations $|\psi_n|^2$ in the case where there are four unstable modes. The parameters are $N=128, ~\Lambda=2.0, ~p=\pi$ and $\xi=10^{-4}$. Lighter shading represents higher site populations. }
\label{f6}
\end{figure}

For a lattice with a large number of sites the one-peak condition is highly impractical because the interaction strength $\Lambda$ needs to be extremely small and the pulse revival period is long. For reasonable interaction strengths, Eq. (\ref{e18}) suggests that there may be more than one unstable mode. Fig. \ref{f6} is a typical representative of the dynamics of the BEC for multiple unstable modes. Here we take $N=128, ~\Lambda=2.0, ~p=\pi$ and $\xi=10^{-4}$. Each bright white spot represents a pulse. As before the right edge of the plot wraps around to the left edge by virtue of the periodic boundary conditions. The random noise seeds approximately four pulses. However, these pulses are not independent of each other. Presumably because of nonlinear mode-mode interaction, they move around, join and split as they collapse and revive.

.

\subsection{\label{sec:level1} Evolution in Fourier space}

The recurrences observed in the peaking events in the DNLSE resemble the energy recurrences in the Fermi-Pasta-Ulam (FPU) problem \cite{Fermi}. The FPU model deals with the evolution of a lattice chain with nonlinear interactions between the nearest-neighbor atoms when  initially a single low-energy mode is excited. For a time scale much longer than the time period of the normal modes, the energy is well localized to the given excited mode, while the amplitudes of the higher-energy modes decay exponentially as a function of the energy difference from the initially excited mode. For a longer time scale it has also been noticed that recurrence of the initial excited mode is possible.

The pulsating behavior of the density distribution of the BEC atoms in the lattice can be viewed as a similar recurrence phenomenon as observed in the FPU model. We have started with a steady state for a given flow quasimomenta ($p>\pi/2$) and a suitable nonlinear interaction strength to  trigger the instability in the system. Ergodicity immediately suggests that the energy initially fed into a single mode should distribute evenly between all Fourier modes. However, the excitation amplitudes of the modes other than the mode corresponding to the initial steady state seems to decay exponentially with the index $Q$. The energy localization to a few Fourier modes in a nonlinear system is not a new phenomenon \cite{Dauxois}. The existence of discrete breathers in a nonlinear lattice system is an example. Recently, energy localization in Fourier space in a so called `q-breather' has been investigated in \cite{Flach}.

\begin{figure}
\includegraphics[clip,width=1.0\linewidth]{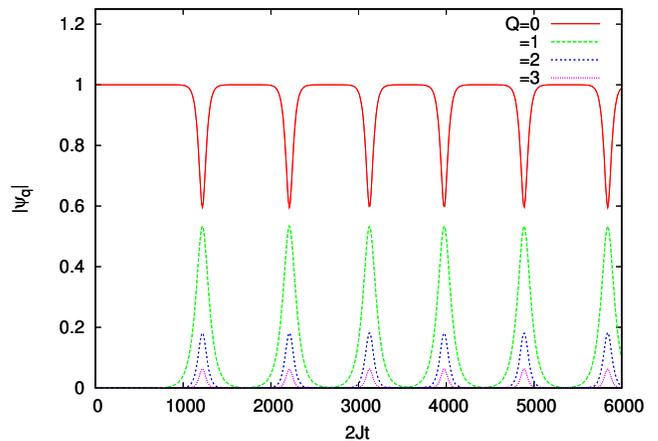}
\caption{Time evolution of the Fourier modes $\psi_q$ of the lattice. The parameters are $N=32$, $p=\pi$ and $\Lambda=0.48$. The mode $Q=0$ corresponds to the initial steady state while modes $Q=1,2,3\dots$ are the low-lying excitations. Only the modes for $Q>0$ are shown.}
\label {f7}
\end{figure}

Figure \ref{f7} shows the time evolution of the Fourier modes $\psi_q$ of DNLSE for the parameters $ \Lambda=0.48,~N=32$. Only a few components are seen to be excited, as the amplitudes of the higher-energy modes are suppressed exponentially. The dominant Fourier components are $P$ and $P\pm1$ which implies that the excitation modes that go unstable will have indices $Q=\pm 1$ with respect to the initial steady state, as expected.

\section{\label{sec:level2}Double Well Analogy}
In order to explain qualitatively the pulsating behavior of the density distribution of the BEC in the lattice, we study a coupled double-well system. Assuming $\psi_{1,2}=|\psi_{1,2}|e^{i\phi_{1,2}}$, the coherent dynamics of such a system can be described by a pair of equations \cite{Raghavan},

\begin{eqnarray}
&&\dot z(t)=-\sqrt{1-z^2(t)}\sin\phi(t), \nonumber\\
&&\dot \phi(t)=\Lambda z(t)+ {z(t)\over \sqrt{1-z^2(t)}} \cos\phi(t),
\label{IV1}
\end{eqnarray}
where $z=|\psi_{2}|^2-|\psi_{1}|^2$ and $\phi=\phi_{2}-\phi_{1}$ are the fractional population imbalance and the relative phase between the two wells. The normalization is $|\psi_{2}|^2+|\psi_{1}|^2=1$. The Hamiltonian (the total energy) in these variables becomes,
\begin{equation}
H={\Lambda z^2\over 2}- \sqrt{1-z^2}~\cos\phi.
\label{IV2}
\end{equation}

\begin{figure}
\includegraphics[clip,width=1.0\linewidth]{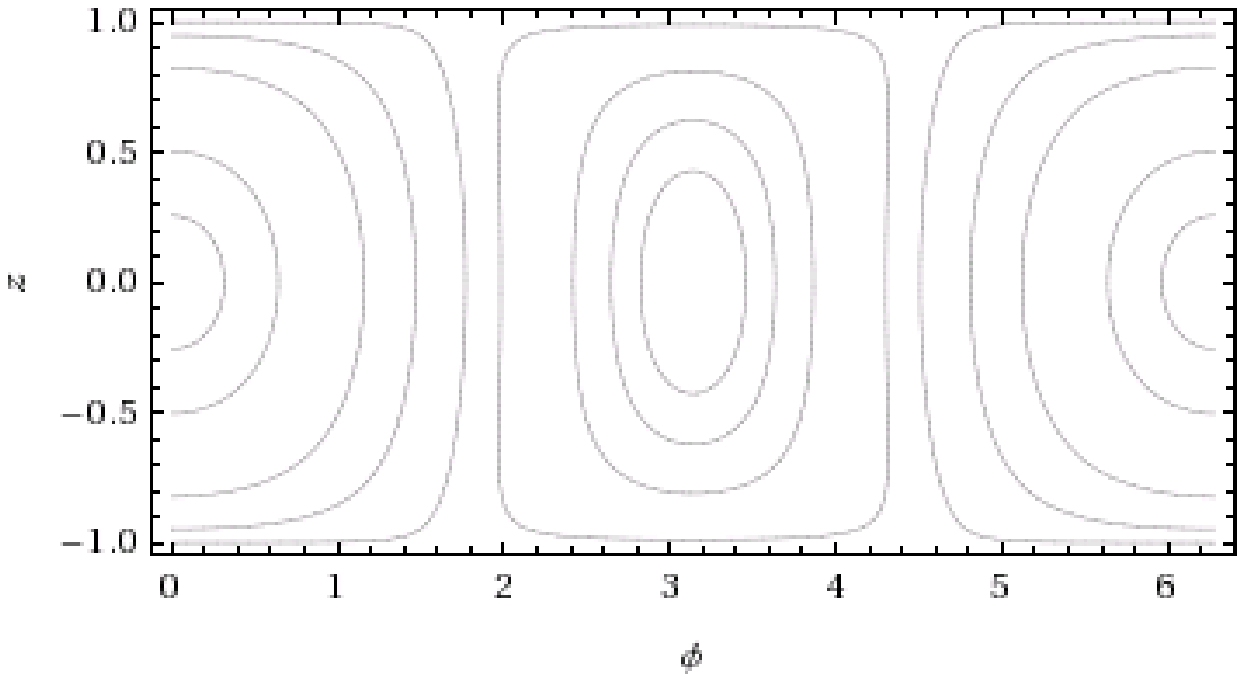}
\includegraphics[clip,width=1.0\linewidth]{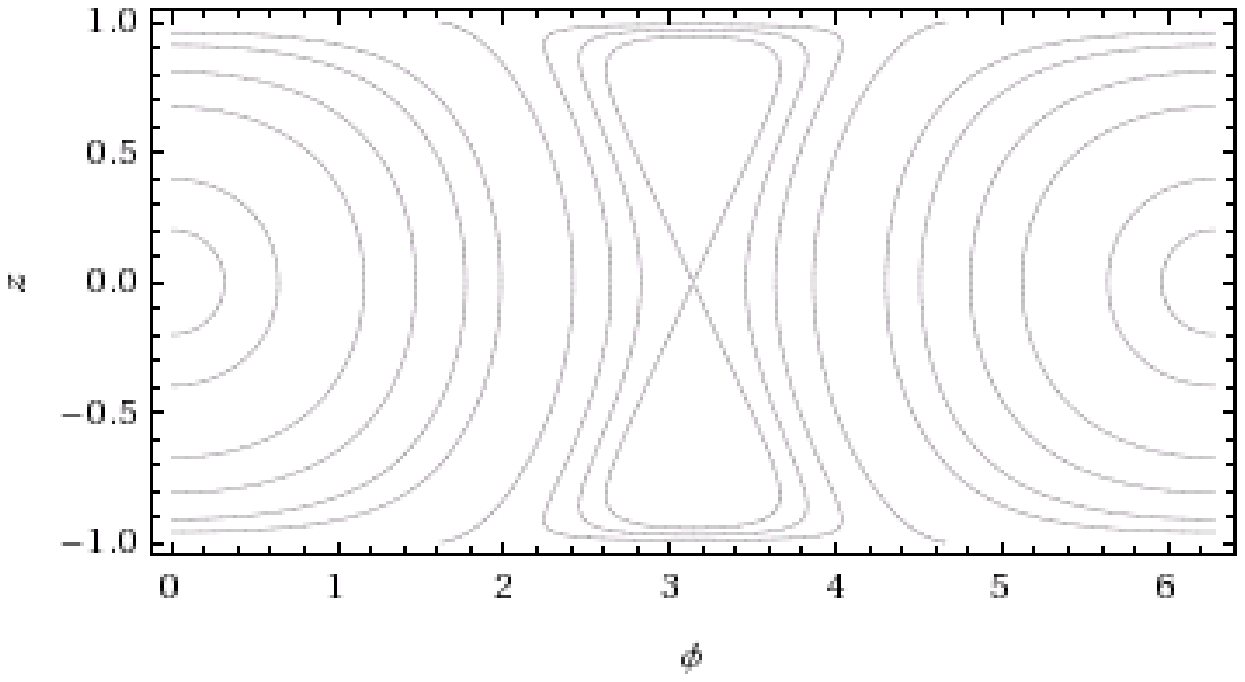}
\caption{Contour plots for the two site Hamiltonian in the $(\phi,z)$ plane for interaction strengths $\Lambda=0.5$ (a) and $1.5$ (b). }
\label{f8}
\end{figure}

Both the norm and the Hamiltonian are the constants of the motion and thus the double-well system , in principle, is integrable. By inspection it can be checked that the fixed points of  Eq. (\ref{IV1}) are $ z=0 , \phi=n \pi$, where $n$ is an integer. The potentially unstable steady state in the multiwell system can be translated into the two-well system by taking the solution $z=0$ and $ \phi=\pi$. The behavior of the orbits near this equilibrium point can be examined by using linear stability analysis as before. It can be easily verified that the state $ \{0,\pi\}$ is stable for the values $\Lambda\leq 1$, and unstable otherwise. In Fig. \ref{f8} we have shown the energy contours of the two-well system for $\Lambda=0.5$ (a) and $1.5$ (b). We have drawn the $\phi$ axis from $0$ to $2\pi$ so that the potentially unstable fixed point $(0,\pi)$ lies at the centers of the plots. For $ \Lambda=0.5$ the potentially unstable steady state is an elliptic fixed point and the time evolution takes the system periodically around this point. At $\Lambda=1$ the elliptic fixed point bifurcates, and for $\Lambda=1.5$ there is a homoclinic orbit with the emergence of two symmetric off-centered elliptic fixed points. Thus, starting in the vicinity of what used to be the potentially unstable steady state, the system takes off in an unstable direction along the homoclinic orbit and goes around one of the bifurcated elliptic fixed points.

The double-well system allows an analytic solution in a closed form in terms of Jacobian elliptic functions. Here we have expressed a solution valid in the range $0\leq z\leq1$,

\begin{eqnarray}
z(t)=\frac{(\text{z}_1-\text{z}_2) \text{z}_4 ~\text{sn}(t \gamma |k)^2+\text{z}_1
   (\text{z}_2-\text{z}_4)}{(\text{z}_1-\text{z}_2) ~\text{sn}(t \gamma
   |k)^2+\text{z}_2-\text{z}_4},
\label{IV3}
\end{eqnarray}
where we have defined
\begin{eqnarray}
&&\text{z}_1=\sqrt{\frac{p+\alpha}{2}},\quad
\text{z}_2=\sqrt{\frac{p-\alpha}{2}},\nonumber\\
&&\text{z}_3=-\sqrt{\frac{p-\alpha}{2}},\quad
\text{z}_4=-\sqrt{\frac{p+\alpha}{2}},\nonumber\\
&&p=\frac{2 a H-1}{a^2},
\quad\alpha=\sqrt{p^2+4q},\nonumber\\
&&q=\frac{1-H^2}{a^2},\quad
a=\frac{\Lambda}{2},\nonumber\\
&&\gamma=a\sqrt{(\text{z}_1-\text{z}_3)(\text{z}_2-\text{z}_4)},\nonumber\\
&&\text{and}\nonumber\\
&&k=\frac{(\text{z}_1-\text{z}_2)(\text{z}_3-\text{z}_4)}{(\text{z}_1-\text{z}_3)(\text{z}_2-\text{z}_4)},\nonumber
\end{eqnarray}
and $H$ stands for the conserved value of the Hamiltonian. The norm which is also a conserved quantity is taken to be equal to one.

\begin{figure}
\includegraphics[clip,width=1.0\linewidth]{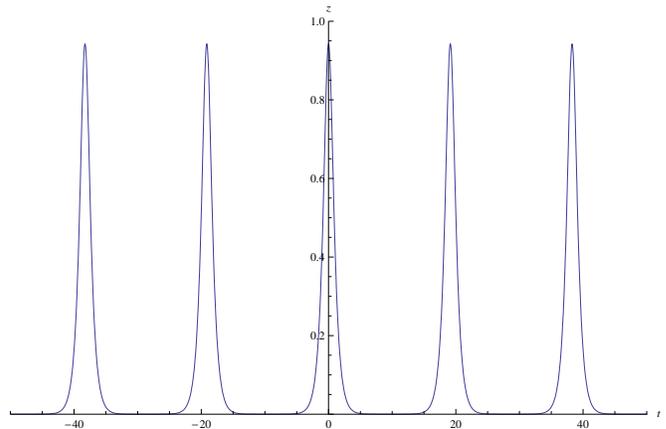}
\caption{Time evolution of the population imbalance $z$ given by the equation Eq. (\ref{IV3}) for the parameter $\Lambda=1.5$ and the Hamiltonian $H=1.0$.}
\label{f10}
\end{figure}

In Fig. \ref{f10} we have plotted the function $z(t)$ given by Eq. (\ref{IV3}) for the parameters $\Lambda=1.5$ and $H=1.0$ such that the double-well system is unstable in linear stability analysis. These parameters correspond to the energy contour close to the homoclinic orbit and bifurcated fixed points (Fig. \ref{f7}(b)). The oscillation in the population imbalance has a striking resemblance to the Fig. \ref{f5} with the plot of the overlap in the lattice system and can be viewed as an analogue of the pulsating instability.

Though we have presented  different versions of the pulsating instability in the two-site system  [ (a) energy contours and (b) the population imbalance ], they describe the same physics. The dynamically unstable system performs periodic oscillation where the system recedes far away from the unstable state and subsequently returns to this state.

The multi site system basically shares the dynamics of the two-site system in a multi-dimensional phase space: Starting from random noise in the neighborhood of an unstable steady state, the system evolves away from, and returns to, the initial state and the process repeats. These periodic recurrences occur in a $2N$-dimensional phase space on the constant energy surface in full analogy with the two-site system. The two-site system is strictly periodic since there is no motion out of the surface; a 1D curve. However, in the multiwell case the dimension of the constant-energy surface is $2N-1$. Our pulsating instability strongly suggests that the system stays close to the homoclinic orbit while it evolves, but depending on the initial noise it still has a large state space to explore. But in the nonlinear multidimensional system the noise may cause the motion to deviate slightly. Upon looping around one of the stable fixed points, the multisite system therefore does not have to return to exactly where it started from. This may account for the slight variations in the period of the pulsations.

\section{\label{sec:level2}Truncated Wigner approximation}

In section II we have discussed the dynamics of a BEC within the classical mean-field theory. The GP equation can in general be very accurate in modeling a weakly interacting BEC. In optical lattices, however, the kinetic energy is represented by the hopping of atoms between adjacent lattice sites. This can be significantly reduced in deep lattices, resulting in enhanced effect of interactions and quantum fluctuations. In the following we include quantum fluctuations in the atom dynamics using stochastic phase space methods. Within the truncated Wigner approximation (TWA) we unravel quantum dynamics into individual stochastic trajectories and calculate expectation values of physical observables by ensemble averaging a large number of trajectories.

For multi-mode dynamics TWA was introduced in nonlinear optics in the studies of quantum fluctuations \cite{DRU93}. Details how to implement TWA in different atomic BEC systems may be found, e.g., in Refs.~\cite{Steel,SIN02,ISE06}.
In the TWA one neglects the third-order derivatives in the generalized Fokker-Planck type equation for the Wigner distribution function \cite{Gardiner}. This allows us to write a nonlinear stochastic differential equation for the Wigner distribution $\psi^W$ of the many-particle wavefunction. For a closed system, this equation is similar to the GP equation with stochastic initial conditions.

Here we apply TWA formalism to quantum atom dynamics in optical lattices. Both zero and finite temperature nonequilirium dynamics has previously been successfully studied in 1D lattice systems in a number of works
\cite{POL03c,ISE05,ISE06,RUO05,Ferris}. The effects of dynamical instabilities in lattices and TWA have been explicitly addressed in Refs.~\cite{POL03c,RUO05,Ferris}. Since quantum fluctuations in an optical lattice can have a notable effect, we pay a special attention to evaluating the correct quantum statistical correlations for the initial state within the Bogoliubov approximation. The emphasis on quantum fluctuations is quite different from typical finite temperature dominated TWA approaches in higher dimensions \cite{BIS08}.

We vary the effective 1D interaction strength $\chi/N_T^2$, for a fixed $\chi$, or a chemical potential \cite{RUO05}. Quantum fluctuations become dominant in the limit of small atom numbers and/or for strong effective 1D interaction strength $g$. In the limit of $N_T\rightarrow\infty$ (for a fixed $\chi$) we recover the classical GP dynamics.

For a closed system, where we ignore any dissipation terms, the TWA dynamics follows from
the stochastic classical field equation, similar to GP equation,
\begin{equation}
i{\partial\over\partial t}\psi_{n}^W=-J(\psi_{n+1}^W+\psi_{n-1}^W)+\chi \abs{\psi_{n}^W}^2\psi_{n}^W\,.
\label{V1}
\end{equation}
The difference from the GP evolution is that we generate a stochastic collection of the initial states and $\psi^W$ is a classical Wigner representation of the full field operator. We evolve each stochastic realization of an initial state accordingly to Eq.~(\ref{V1}) and evaluate corresponding ensemble averages. Our TWA formalism is very similar to the one used in Refs.~\cite{ISE05,ISE06,RUO05}, except that in each TWA realization we fix the total atom number \cite{MAR08}.

\subsection{\label{sec:level2}Initial State}

In order to generate the initial state stochastically  within TWA, we solve the quasiparticle excitation spectrum using the Bogoliubov approximation. We again consider a stationary solution for a moving plane wave $\phi_{0n}=\text e^{i(p n-\omega(p)t)}$, with $\omega(p)=-2J\cos p+\chi$. The linearized fluctuations around the stationary solution are obtained from
\begin{equation}
\hat{\psi}_{n}(t)=\phi_{0n} \hat\alpha_0 +\delta \hat{\psi}_n(t)\,,
\label{V11}
\end{equation}
such that the total number of condensate particles
\begin{eqnarray*}
N_c=\langle\hat\alpha_0^\dagger \hat\alpha_0\rangle,
\end{eqnarray*}
which is much larger than one. Analogously to our earlier classical Bogoliubov treatment, the fluctuation part, $\delta\hat{\psi}_{n}(t)$, can be written, in terms of quasiparticle operators $\hat{\alpha_q},\hat{\alpha_q}^\dagger$, as
\begin{equation}
\delta\hat{\psi}_{n}(t)=\sum_q (u_q\hat{\alpha_q} \text e^{i(n q -\Omega_q t)}+v_q^*\hat{\alpha_q}^\dagger\text e^{-i(n q -\Omega_q ^*t)}).
\label{V12}
\end{equation}
The operators $\hat{\alpha_q},\hat{\alpha_q}^\dagger$ obey Bose commutation relations, $[\hat{\alpha_q},\hat{\alpha_{q'}}^\dagger]=\delta_{q,q'}$. Here the normal mode frequency $\Omega_q$, and the quasiparticle amplitudes $u_q$, and $v_q$ are given by Eqs.~(\ref{e15}), (\ref{ee18}), and~(\ref{ee19}). The quasimomentum is denoted by $q$.

The total number of non-condensate particles in the Bogoliubov theory is given by
\begin{equation}
N_{nc}=\sum_{n,q}(|u_n^q|^2+|v_n^q|^2) \langle\hat\alpha_q^\dagger \hat\alpha_q \rangle+\sum_{n,q}|v_n^q|^2,
\label{V13}
\end{equation}
with
\begin{equation}
\langle \hat\alpha_q^\dagger \hat \alpha_q \rangle \equiv \bar n=[e^{\Omega_q/{K_B T}}-1]^{-1}.
\label{V14}
\end{equation}
At $T=0$, $\langle \hat\alpha_q^\dagger \hat \alpha_q \rangle=0$, and the non-condensate fraction is simply obtained from
\begin{equation}
N_{nc}=\sum_{n,q}|v_n^q|^2.
\label{V15}
\end{equation}

In order to construct the initial state within TWA we replace the quantum operators $(\hat{\alpha_q},\hat{\alpha_{q'}}^\dagger) $ in Eq. (\ref{V12}) by complex stochastic variables $(\alpha_q,\alpha_{q'}^*)$ obtained by sampling the corresponding Wigner distribution function. Our formalism
follows Ref.~\cite{ISE06}, except that here we fix the total atom number, so that in the TWA simulations the condensate and the non-condensate atom number fluctuations are related \cite{MAR08}. In the Bogoliubov approximation the operators $\alpha_q,\alpha_q^\dagger$ behave as a collection of ideal harmonic oscillators. The Wigner function at $T=0$ reads \cite{Gardiner}
\begin{equation}
W(\alpha_q,\alpha_q^*)=\frac{2}{\pi}\exp[-2|\alpha_q|^2]\,.
\label{V16}
\end{equation}
	The function $W(\alpha_q,\alpha_q^*)$ is a Gaussian with the width 1/2. Here the nonzero width mimicks the quantum noise. Due to the nonzero width of the vacuum modes in the Wigner distribution, each unoccupied phonon mode begins with uncorrelated Gaussian noise, distributed over the plane wave basis, and normalized to an average of a half particle per mode. This provides a seeding for scattering events in the dynamics, but in the end it is subtracted out from all normally-ordered quantum averages. For each stochastic realization, the number of non-condensate atoms reads
\begin{equation}
N_{nc}=\sum_{n,q}(|u_n^q|^2+|v_n^q|^2) (\alpha_q^* \alpha_q-\frac{1}{2})+\sum_{n,q}|v_n^q|^2,
\label{V17}
\end{equation}
which may fluctuate about the mean value $\sum_{n,q}|v_n^q|^2$. The ensemble average over many realizations is $\langle \alpha_q^* \alpha_q\rangle_W=\frac{1}{2}$. Since the total particle number $N_T$ is conserved, the number of condensate atoms in each individual run is given by
\begin{equation}
N_{c}=N_{T}-N_{nc}.
\label{V18}
\end{equation}
Finally, we set
\begin{equation}
\alpha_{0}=\sqrt{ N_c +\frac{1}{2}}
\label{V19}
\end{equation}
in the initial state Eq. (\ref{V11}). Note that, even though we consider a uniform system with a plane wave phonon basis and uncorrelated noise in the initial phonon modes, the fixing of the total atom number introduces long wavelength correlations in the system between the condensate mode and the excited quasiparticle modes \cite{MAR08}.

\subsection{\label{sec:level2}Numerical Realization}

\begin{figure}
\includegraphics[clip,width=1.0\linewidth]{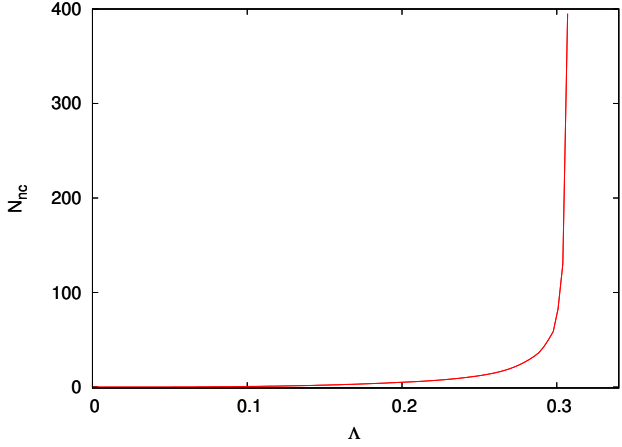}
\caption{Number of non-condensate atoms given by Eq. (\ref{V15}) as a function of the scaled interaction strength $\Lambda$ within Bogoliubov approximation. Note that the critical interaction strength for the onset of the instability is 3.08. For the interacting initial state in the simulations we consider $\Lambda=0.284$ that corresponds to $N_{nc}\simeq 30$.}
\label{fp10}
\end{figure}

We study the non-equilibrium quantum dynamics of a BEC within TWA. We consider a BEC in a lattice initially in a stable steady state which is crucial for the validity of the TWA. At the beginning of the time evolution the lattice is driven to a dynamically unstable regime, for instance, by accelerating it through $p=\pi/2$ or by modifying the atom-atom interactions. Here in our simulations, we fix the initial velocity and change the value of the atom-atom interactions. We consider a small or a zero depletion of atoms from the condensate in the initial state, so that the Bogoliubov approximation is valid. For the case of a non-interacting initial state the average number of non-condensate atoms is zero. We set the initial momentum to be $\pi$. As an interacting initial state, we consider in all our simulations (unless otherwise stated) the rescaled interaction strengths, $\Lambda(\chi/2J)=0.284$ 
corresponding to the average non-condensate atom number $N_{nc}\simeq 30$. It should be noted that the critical value of $\Lambda$ for the onset of the instability is 0.308. The number of lattice sites $N$ is always taken to be 32, though larger lattices can also be simulated. In all the simulations we vary the total atom number and the interaction strength $g\propto\chi/N_T$, so that $\chi$ (and chemical potential) remains constant. Then the ratio $\chi/N_T^2$ represents the effective strength of the interactions in the system \cite{RUO05}. The atom-atom interactions are turned up instantaneously to a desirable value so that the system evolves in the classical dynamically unstable regime.  
In all time evolutions we take $\Lambda=0.48$. Choosing the initial state closer to the onset of the dynamical instability would have resulted in a larger depletion of atoms from the condensate, as the non-condensate atom number in the Bogoliubov theory diverges at the instability threshold (see Fig. \ref{fp10}). 

For each individual realization of the time evolution of the ensemble of the Wigner distributed  wave functions we sample the initial state according to the previous section. The generation of the initial state consists of replacing the operators $(\hat\alpha,\hat\alpha^\dag)$ by complex, Gaussian distributed variables $(\alpha,\alpha^*)$. We have used the Box-Mueller algorithm \cite{NR} for the sampling. As before we integrate the dynamical equation (Eq. (\ref{V1})) using the FFT split-step method \cite{Javanainen}.

\subsection{\label{sec:level2}Results}

Since the TWA returns symmetrically ordered expectation values, instead of normally ordered ones, we need to calculate the normally ordered expectation values from the simulation data \cite{ISE05,ISE06}. Here we are only considering the lowest energy band in the tight-binding approximation, so normally ordering the operator expectation values is straightforward. According to Ref.~\cite{ISE06}, we have (here $\bold x$ always refers to one given site) the atom number in a lattice site
\begin{equation}
n(\bold x)=\langle\psi^*(\bold x)\psi(\bold x)\rangle_W -\frac{1}{2}\,,
\end{equation}
with the corresponding fluctuations
\begin{equation}
\Delta n(\bold x)=\sqrt{\langle(\psi^*(\bold x)\psi(\bold x))^2\rangle_W-\langle\psi^*(\bold x)\psi(\bold x)\rangle_W^2-\frac{1}{4}}\,.
\end{equation}
The normalized phase coherence along the lattice follows from \cite{ISE05,ISE06}
\begin{equation}
C(\bold x,\bold y)=\frac{\langle\psi^*(\bold x)\psi(\bold y)\rangle_W}{\sqrt{n(\bold x)n(\bold y)}}.
\end{equation}
The overlap of the field amplitudes between times $t=0$ and $t=\tau$, which is a measure of the revival of the pulse, is given by
\begin{equation}
f^W(\tau)= \langle \sum_n\psi_n^*(\tau)\psi_n(0)\rangle_W.
\end{equation}


In Fig.~\ref{f10} we show a typical single-trajectory result for the overlaps of the state of the system with the initial state as a function of time for an interacting initial state with $\Lambda=0.284$ that corresponds to the number of non-condensate atoms $N_{nc}\simeq 30$, and quasimomentum $p=\pi$, for the various values of the total number of atoms (a) $N_T=10^6$, (b) $N_T=10^4$, (c) $N_T=10^3$, and (d) $N_T=500$. 
As stated earlier, we instantaneously turn the interaction on to the value $\Lambda=0.48$ so that the system evolves in the dynamically unstable regime. We vary the atom-atom interactions $g$ and the atom number $N_T$ but keep the value of $\chi ~(\text{or} ~\Lambda) $ fixed for each simulations.  

For a smaller total number of atoms, the non-condensate atom fraction in the initial state and the scattering length are larger and, consequently, quantum effects are generally more observable. Each plot clearly indicates the pulsating instability without any noticeable damping. In each case the non-condensate particles simply act as a vacuum noise in the system. Note that the pulsating period depends on the number of particles; the smaller the number of particles the shorter the period of oscillation. This can be qualitatively understood in terms of the scattering events between condensate and non-condensate particles. The rate of scattering processes depends on the number of non-condensate particles. Higher scattering rate leads to faster condensate depletion, and thus shorter period of the oscillation.

\begin{figure}
\includegraphics[clip,width=1.0\linewidth]{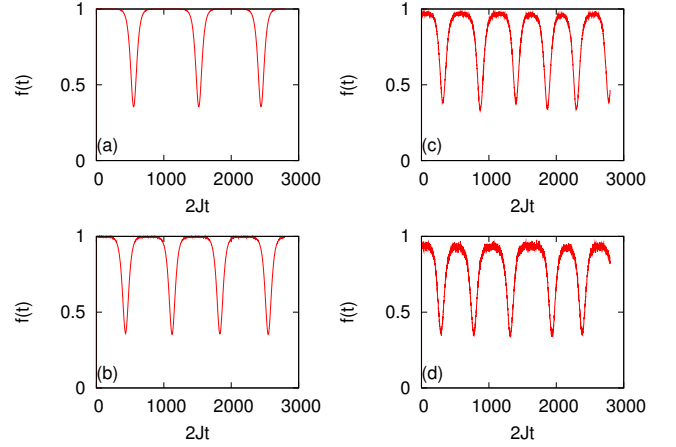}
\caption{Overlap of a single realization as a function of time sampled from the Wigner distribution for various numbers of particles (a) $N_T=10^6$, (b) $N_T=10^4$, (c) $N_T=10^3$ and (d) $N_T=500$. The initial state is interacting with the non-condensate atoms $N_{nc}=30$. The parameters of the simulation are $\Lambda=0.48, ~N=32$ and $p=\pi$. There is no significant damping in the oscillation even if the non-condensate noise is substantial.}
\label{f10}
\end{figure}

\begin{figure}
\includegraphics[clip,width=1.0\linewidth]{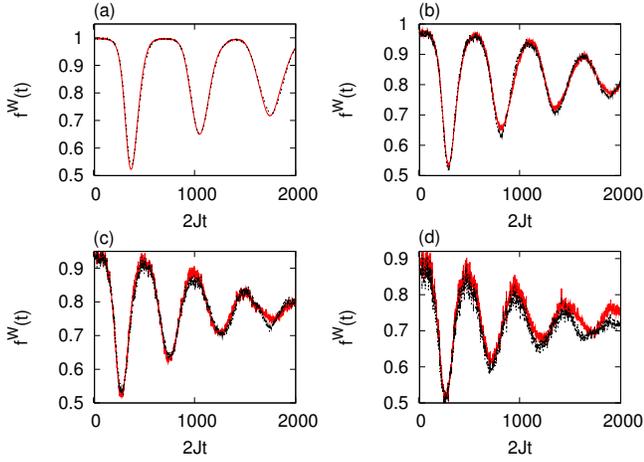}
\caption{Comparison of the ensemble average of the overlap of the state of the lattice sampled over $400$ realizations for the initial number of non-condensate particles $N_{nc}=0$ and $N_{nc}=30$, and for the total number of particles (a) $N_T=10^4$, (b) $N_T=10^3$, (c) $N_T=500$ and (d) $N_T=300$. The parameters of the simulations are $\Lambda=0.48, ~N=32$ and $p=\pi$. There is no significant change in the nature of time evolution in two different initial state when the total atom number is large. However, the curves start deviating as the number of particles are reduced.}
\label{f21}
\end{figure}

\begin{figure}
\includegraphics[clip,width=1.0\linewidth]{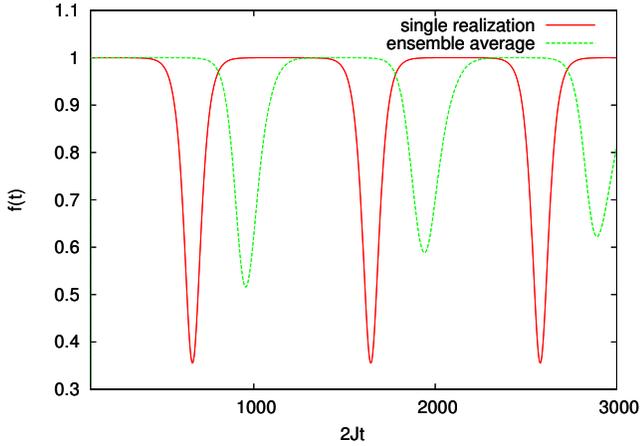}
\caption{Comparison of the overlaps in a typical single realization versus an ensemble average sampled over 400 realizations with the initial interacting state for $N_{nc}=30$ and $N_T=10^6$. The parameters of the simulations are $\Lambda=0.48, ~N=32$ and $p=\pi$.}
\label{f12}
\end{figure}

\begin{figure}
\includegraphics[clip,width=1.0\linewidth]{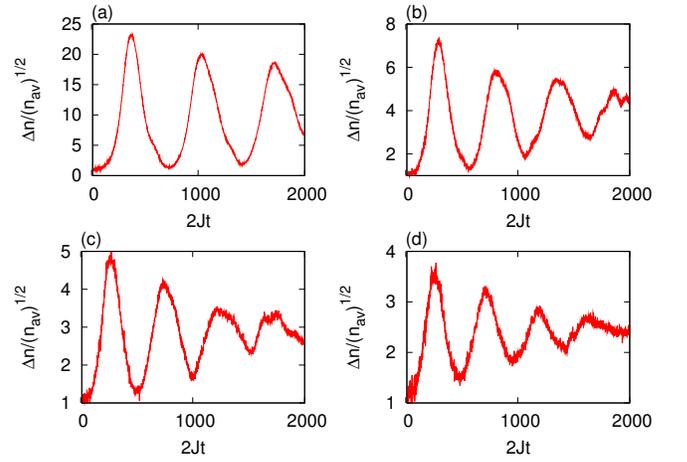}
\caption{Ensemble average of the relative number fluctuations sampled over $400$ realizations in the central lattice site as a function of time for the total numbers of paricles (a) $N_T=10^4$, (b) $N_T=10^3$, (c) $N_T=500$ and (d) $N_T=300$ with $N_{nc}=30$. The parameters of the simulations are $\Lambda=0.48, ~N=32$ and $p=\pi$.}
\label{f13}
\end{figure}	


Figure \ref{f21} represents an ensemble average of the overlap $f^W(t)$ sampled over 400 trajectories for the total number of atoms (a) $N_T=10^4$, (b) $N_T=10^3$, (c) $N_T=500$ and (d) $N_T=300$ for interacting (full) and non-interacting (dash) initial state. For the non-interacting initial state the non-condensate atom number is zero whereas for the interacting initial state we again take $\Lambda=0.284$ that corresponds to the non-condensate atom number $N_{nc}\simeq30$. For each simulations, both for interacting and non-interacting initial state, the time evolution is carried out by varying the $N_T$ and $g$ and instantaneously switching the interactions to a value such that $\Lambda=0.48$. The other parameters of the simulations are $N=32$ and $p=\pi$. Though the sampling noise is still there, especially for smaller atom numbers, the figures clearly show a damping in the pulsation which contrasts the time evolution in the single realization. The figure also shows that the smaller the atom numbers, the higher the damping rate. Moreover, there is no noticeable difference in the time dynamics for the two different initial states as long as the atom numbers are large. However, there is a significant deviation in the time dynamics for these two different initial states for smaller numbers atoms. Fig. \ref{f12}  represents a comparison of the overlaps in a typical single realization versus that in an ensemble average for the larger atom number $N_T=10^6$, and with the interacting initial state. The other parameters of the simulations are same as in Fig. (\ref{f21}). 

Similar sort of damping is also observed in the number fluctuations at a given site. In Fig. \ref{f13} we show ensemble averages of the relative the number fluctuations $\Delta n/n_{av}$ at the central lattice site for the total numbers of the atoms (a) $N_T=10^4$, (b) $N_T=10^3$, (c) $N_T=500$ and (d) $N_T=300$ for an interacting initial state. All the parameters of the simulations including the initial state are the same as in Fig. (\ref{f21}). In addition to the damping in the oscillations the figure also reveals that the mean relative number fluctuations are larger as the number of atoms gets small.

Our main focus in this work is to study the inherent quantum effects on the pulsation phenomenon of the BEC in an optical lattice. For that purpose it is obvious to look into the various uncertainties associated with the pulse. For instance we calculate the amplitude uncertainty of the first pulse in the pulsating instability as a function of the total number of atoms

\begin{equation}
\Delta({\bold A})=\sqrt{\langle A^2\rangle_W-\langle A\rangle^2_W}.
\end{equation}

\begin{figure}
\includegraphics[clip,width=1.0\linewidth]{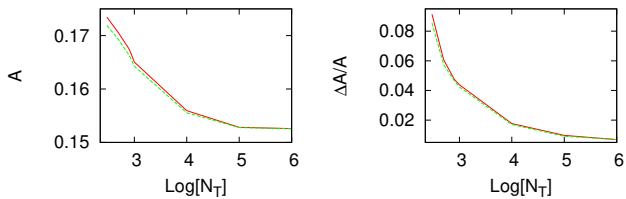}
\caption{Ensemble average of the amplitude (left) of the first peak and the corresponding uncertainty (right) during pulsation as a function of the total number of atoms, $N_T$ for interacting (full) and non-interacting (dash) initial states. The average is taken over 400 realizations. The parameters of the simulation are same as in Fig. \ref{f21}. The amplitudes remain unaffected for higher atom numbers, and the fluctuations acts as a simple random noise to initiate the instability. However, the quantum noise manifests itself, both in the amplitude and its uncertainty as the number of atoms gets small.}
\label{f14}
\end{figure}

Fig. \ref{f14} represensts the amplitudes of the first pulse and the corresponding  uncertainties  sampled over 400 realizations as a function of the total number of atoms, both for the interacting (full) and the non-interacting (dash) initial states. The parameters of the simulations are the same as in Fig. \ref{f21}. For the non-interacting initial state the number of non-condensate atoms is zero. In the interacting initial state we consider the interaction parameter $\Lambda=0.284$ that corresponds the number of non-condensate atoms $N_{nc}\simeq 30$. The figure shows that the uncertainty and the value of the pulse amplitude saturate for the higher atom numbers for both interacting and non-interacting initial states, but rise sharply when the atom number becomes small. This implies the fluctuation dominance at the smaller atom numbers in which the effective interaction is large. The figure also shows that the distinction between these two initial states is apparent only for smaller atom numbers. 

We see that the quantum fluctuations have an effect on the collapse and revival of the pulse. In the case of a single realization the revival seems to be very robust and repeats practically forever. However, the ensemble average over many stochastic realizations in the Wigner method produces a damping in the pulsation. This damping may be due to the decoherence as the number of the non-condensate atoms grows and their interactions with the atoms in the condensate mode increases. The quantum effects are more dominant and the revival of the pulse becomes progressively weaker as the number of the atoms becomes small.
 	
\section{\label{sec:level2}Concluding Remarks}

In this paper we have presented new insights into the unstable dynamics of the BEC in an optical lattice in the limit of weak atom-atom interactions and by incorporating quantum fluctuations. The common belief is that the flow of the dynamically unstable BEC in an optical lattice would be erratic, or lead to the formation of stable solitons. Here we moved a step further and show that, in the classical mean-field theory, the instability may also trigger a quasi-periodic pulsation in the atom density distribution if the atom-atom interactions is weak. The requirement that linear stability analysis finds a single unstable mode gives the scale for the  `weak' nonlinearity and the ensuing pulsating phenomena. 

A qualitative argument has been put forward to explain the pulsating behavior of the dynamics by comparing the lattice system with the integrable double-well system. In the case of two wells the unstable mode leads to a non-trivial dynamics in the population imbalance such that an infinitesimal noise could produce a large-amplitude collective oscillation of the atoms between the wells. An analogous phenomenon is observed in a lattice in the limit of weak atom-atom interactions. We, therefore, surmise that the pulsating instability is a remnant of the integrability.

We incorporate the quantum fluctuations using stochastic phase-space methods. We use the Bogoliubov approximation to generate the initial state for the time evolution of the system. A sequence of the stochastic fields obtained in this way are then used to calculate the expectation values of the observables. We then compare the single realization results with the ensemble averages. It is observed that the quasiperiodic behavior in the time dynamics can still be seen in the single realizations. However, the quantum averages show that the revival of the pulse becomes weaker and weaker as the atom number gets small. 

For experimental realizations, the flow states $p\approx \pi$ near the Brillouin zone boundary can be prepared by accelerating the lattice \cite{FAL04}. Alternatively, by exploiting the symmetry of the DNLSE, every solution $\psi_n(t)$ for the given interaction parameter $\Lambda$ there is a solution $(-1)^n\psi_n ^*(t)$ for $-\Lambda$. That means the state for $p=\pi$ in the repulsive case is equivalent to the state for $p=0$ in the attractive case. This symmetry has already been used to generate solitons in a nonlattice gas \cite{Strecker}. We speculate that the same technique can be used to observe the pulsating instability in the lattice. However, given that the pulsating phenomenon only results in the weak nonlinearity limit, the corresponding time scales for the pulsation can be very long and pose a severe technical challenge.

What would be seen in an experiment depends on how the experiment is carried out. If averaging over repeated experiments is called for, the ensemble averages are the proper quantities to compare with. On the other hand, it might be possible to monitor atom numbers in the lattice continuously, e.g., by off-resonant light scattering. A single realization is therefore observable as a matter of principles. However, the TWA scheme does not take into account back-action of the measurements, which could be severe. This problem area will be the subject of future work.

\end{document}